\documentclass[12pt]{article}

\usepackage{latexsym,amsfonts,amsmath}

\renewcommand{\theequation}{\thesection.\arabic{equation}}
\newcommand{\newsection}{    
\setcounter{equation}{0}\section}

\def\be{\begin{equation}}
\def\ee{\end{equation}}
\def\bd{\begin{displaymath}}
\def\ed{\end{displaymath}}
\def\alppr{{\alpha^\prime}}
\def\cp{$\mathbb{CP}^2$}
\def\Gs{\ensuremath{\not\!\!G}}
\def\Fs{\ensuremath{\not\!\!F}}
\def\lp{\left(}
\def\rp{\right)}
\def\ba{\begin{eqnarray}}
\def\ea{\end{eqnarray}}
\def\Gp{\ensuremath{\lp\Gamma^5+i\Gamma^6\rp\lp\Gamma^7+i\Gamma^8\rp}}
\def\Gm{\ensuremath{\lp\Gamma^5-i\Gamma^6\rp\lp\Gamma^7-i\Gamma^8\rp}}
\def\p{p^+}
\def\sech{\ensuremath{{\mathrm {sech}}}}
\def\vac{\ensuremath{|0\rangle}}
\def\sss{\ensuremath{\scriptscriptstyle}}

\begin{document}
\font\cmss=cmss10 \font\cmsss=cmss10 at 7pt
\vspace{18pt}
\par\hfill Bicocca-FT-02-15
\vskip .1in \hfill hep-th/0207221

\begin{center}
{\LARGE \textbf{Penrose limit of a non--supersymmetric\\ \vspace{4pt}
RG fixed point}}
\end{center}

\vspace{30pt}

\begin{center}
{\textsl{Roberto Casero}}

\vspace{20pt}

\textit{Dipartimento di Fisica, Universit\`{a} di Milano-Bicocca\\
Piazza della Scienza, 3\\
I-20126 Milano, Italy.}

\end{center}

\vspace{30pt}

\begin{center}
\textbf{Abstract }
\end{center}

\vspace{4pt} {\small \noindent We extend the BMN duality between IIB
superstring theory on a pp--wave background and a sector of ${\cal N}
=4 $ super Yang-Mills theory to the non--supersymmetric and unstable
background built by Romans as a compactification on a $U(1)$ bundle
over \cp $\,$with 3--form and 5--form field strength fluxes. We obtain
a stable theory with the fewest number of supercharges (e.g. 16)
allowed by this kind of solutions and make conjectures on the dual
gauge theory.}  \vfill
\vskip 5.mm
 \hrule width 5.cm
\vskip 2.mm
{\small
\noindent
roberto.casero@mib.infn.it}

\eject

\newsection{Introduction} A duality between type IIB strings on the
maximally supersymmetric pp--wave \cite{Gueven} \cite{Blau} and a
sector of ${\cal N} =4 $ super Yang-Mills theory has been recently
proposed. In \cite{BMN} Berenstein, Maldacena, and Nastase show how
string theory on the maximally supersymmetric pp--wave can be obtained
from $AdS_5 \times S^5$ as the Penrose limit \cite{Penrose} along a
null geodesic.

This background is particularly interesting because the superstring
theory can be solved on it using the GS superstring formalism and
light--cone gauge \cite{Metsaev}, \cite{MT}. All the string states are
massive \begin{equation} \label{BMNstr}
H=-p_+=\sum_{n=-\infty}^{+\infty}\: N_n \sqrt{\mu^2+\frac{n^2}{(\alppr
p^+)^2}} \end{equation} where $N_n$ is the occupation number of the
$\mathrm{n^{th}}$ normal mode of the bosonic and fermionic
fields. Choosing a null geodesic breaks $SU(4)_R$ to the $U(1)_R$ of
rotations along the geodesic. BMN argued that string states on the
pp--wave limit of $AdS_5\times S^5$ are dual to a sector of ${\cal N}
=4 $ SYM which is composed of both chiral and non--chiral operators
with large dimension $\Delta$, large $U(1)_R$ charge $J$ and fixed
$\Delta - J$ \begin{equation} \begin{split} \label{ref:BMN} \Delta-J
&= -\frac{p_+}{\mu}=\sum_{n=-\infty}^{+\infty}\: N_n
\sqrt{1+\frac{4\pi g_s N n^2}{J^2}} \\ \frac{\Delta+J}{R^2} &= -\mu\,
p_-
\end{split} \end{equation}
where $R^4 \equiv 4 \pi \alppr^2 g_s N$ is the common radius of
$AdS_5$ and $S^5$ and the expressions are valid for
$\frac{\Delta-J}{J}\ll 1$.

BMN explicitly identified the non--chiral operators corresponding to the 
string spectrum (\ref{BMNstr}). 
Expression (\ref{ref:BMN}) can be expanded as a perturbation series in 
the constant $\frac{g_sN}{J^2}=\frac{\lambda}{J^2}$, which can be 
interpreted as an effective coupling constant. 
In this way, perturbative calculations may be performed in the 
non--perturbative regime $\lambda\gg 1$, and comparison with 
the superstring results can be carried out.

After \cite{BMN} the Penrose limit has been applied to other, less
symmetric, models, and some results were obtained on the
non--perturbative behaviour of a large variety of gauge theories. In
this paper we go in this same direction and consider the large
$\Delta$ and {\it J} (along with large {\it N} and $\lambda$) limit of
a non--supersymmetric unstable gauge theory which is obtained as the
IR fixed point of the renormalization group flow from ${\cal N}=4$ SYM
deformed through a mass term for one of the fermions in the adjoint of
the gauge group \cite{GPPZ} \cite{DZ} \cite{Distler}. The dual to this
theory is the compactification built in \cite{Romans} as a $U(1)$
fibration over \cp. The background has three-- and five--form field
strengths turned on, and is unstable.

After taking the Penrose limit, we find a pp--wave background with
constant NS--NS and R--R field strengths. pp--wave backgrounds with
2--form fields have been considered in \cite{Warner}, \cite{GZS}, and
\cite{CVJ} for a different and supersymmetric fixed--point. Despite
the original instability, our solution is stable and has the minimal
number of supersymmetries allowed for IIB pp--waves (e.g. 16). This has
some interesting consequences. First of all, the bosons and fermions
have different masses, which gives the theory a non--vanishing, finite
and positive zero--point energy. Moreover, in the perturbative
expansion of the anomalous dimension of the dual operators, we find
that some scalars have as the first order correction a term
proportional to $\sqrt{\lambda_{\sss eff}}$ \be
\lp\Delta-J\rp_n=\sqrt{2}\lp 1\!\pm n{\textstyle
\sqrt{\frac{\pi}{8}}}\sqrt{\lambda_{\sss
eff}}+\frac{\pi}{16}n^2\lambda_{\sss eff}+O(\lambda_{\sss
eff}^{3/2})\rp \end{equation} where $\lambda_{\sss eff}=\frac{g_{\sss
YM}^2N}{J^2}$.

We also find that the perturbation series for some fermionic states
starts from the second order \be
\lp\Delta-J\rp_n=\frac{1}{\sqrt{2}}\lp
1+\frac{\pi^2}{16}\,n^4\,\lambda_{\sss eff}^2 +O(\lambda_{\sss
eff}^3)\rp \end{equation} as was also found for a different model in 
\cite{CVJ}.

This paper is organized as follows. In section 2 we review the
construction of the supergravity solution of \cite{Romans}, first
following the original paper, and then in a different way that makes
the symmetries more evident. In section 3 we take the Penrose limit of
this solution. In section 4 we quantize the string theory on the
constant NS--NS and R--R fields pp--wave, and give an approximation to
the zero--point energy. In the last section we discuss the field
theory dual to the pp--wave string theory.


\newsection{The $SU(3)\times U(1)$ supergravity solution} The
2--dimensional complex projective space \cp $\,$is defined as the
subset of $\: \mathbb{C}^3 -\{0\}$ with the identification
$(z_1,z_2,z_3)\simeq (\lambda z_1,\lambda z_2,\lambda z_3)$ with
$\lambda$ any complex number different from zero. The metric of this
space is given by \bd ds^2=\frac{dr^2+r^2 \sigma_3^2}{(1+\frac{\Lambda
\, r^2}{6})^2}+\frac{r^2(\sigma_1^2+\sigma_2^2)}{1+\frac{\Lambda \,
r^2}{6}} \ed where $\Lambda$ is the cosmological constant and
$\sigma_i$, $i=1,2,3$ are the $SU(2)$--invariant forms satisfying
$d\sigma_1=2 \sigma_2 \wedge \sigma_3$ and permutations.  It is
possible to define a connection $A$ that satisfies 
\begin{equation} \label{killcp}
d\eta+(\frac{1}{4}\omega_{ab}\Gamma^{ab}-i\,e\,A)\eta=0 \end{equation} 
where {\it
e} is the charge of the spinor $\eta$ (we use the same charge
normalization as \cite{Romans}): \bd A=A_{\mu}dx^{\mu}=3 \:
\frac{C_{\mu\nu}x^{\nu}dx^{\mu}}{1+x_1^2+x_2^2+x_3^2+x_4^2} \qquad ,
\qquad C=\left( \begin{array}{cccc} 0 & 1 & 0 & 0 \\ -1 & 0 & 0 & 0 \\
0 & 0 & 0 & 1 \\ 0 & 0 & -1 & 0 \\
\end{array} \right)
\ed and $z_1=x_1+i\,x_2$, $z_2=x_3+i\,x_4$ and
$z_1^2+z_2^2+z_3^2=\frac{6}{\Lambda}$.  The solution to equation
(\ref{killcp}) is given by \begin{equation} \label{solcp} \left\{
\begin{array}{llll} e=\frac{1}{2} \\ (\Gamma_{12}-\Gamma_{34})\eta=0
\\ (i+\Gamma_{12})\eta=0 \\ \partial_i\eta=0 \quad i=1,2,3,4
\end{array} \right.  \end{equation} 
and imposing $\eta^{\dagger} \eta =1$, $\eta$
can be determined up to a phase $\alpha$. If we define the charge
conjugate spinor $\chi \equiv C \eta^\ast$, we can build a complex
(2,0)--form $K\equiv\chi^\dagger\,\Gamma_{ij}\,\eta\, e^i \wedge e^j$
($\{e^i\}$ $(i=1,\ldots ,4)$ are the vielbeins of \cp) which will be
useful later to define the complex three--form on the 10--dimensional
background. Given (\ref{solcp}) we find that \begin{equation} \left\{
\begin{array}{ll} K_{12}=K_{34}=0 \\
K_{13}=-K_{24}=iK_{23}=iK_{14}=e^{i\alpha} \end{array} \right.
\end{equation} where $\alpha$ is an arbitrary real constant.

The connection {\it A} can be used to build a $U(1)$ bundle over
\cp. We refer to this space as $M_5$. Its metric is given by
\begin{equation} ds^2_{M_5}=ds^2_{\mathbb{CP}^2} + c^2 (d\tau -A)^2
\end{equation} for a constant {\it c} to be determined later.

Given $M_5$ we can build a compactification of IIB supergravity of the
form $AdS_5 \times M_5$ \cite{Romans} \begin{equation} \label{ansatz}
\begin{split}
ds^2&=ds^2_{AdS_5}+ds^2_{M_5} \\ G_3&= \frac{2}{R} \,e^{-i \tau}
K_{ij} \, e^i\wedge e^j \wedge e^5 \\
F_5&=-\frac{1}{\sqrt{2}R}\,(e^\mu \wedge e^\nu \wedge e^\rho \wedge
e^\sigma \wedge e^\tau+e^m\wedge e^n\wedge e^p\wedge e^q\wedge e^r)
\end{split} \end{equation} 
where $\{e^\mu\}$ and $\{e^m\}$ are the vielbeins of
$AdS_5$ and $M_5$ respectively, and
$\Lambda_{\mathbb{CP}^2}=\frac{1}{c^2}=\frac{8}{R^2}$.

We choose a coordinate system in which the $AdS$ metric is given by
\bd ds^2_{\mathrm{AdS}}=R^2(-\cosh^2 \!\!\rho \;dt^2 + d \rho^2
+\sinh^2 \!\!\rho \;d \Omega_3 ) \ed and the complex coordinates of
\cp are \begin{equation} \label{parcp}
\begin{array}{ll} 
z_1=\cot \omega \cos\frac{\theta}{2}e^{i\frac{\psi+\phi}{2}} 
\\ z_2=\cot \omega \sin\frac{\theta}{2}e^{i\frac{\psi-\phi}{2}} 
\end{array}
\end{equation} The ten--dimensional metric takes the form
{\setlength\arraycolsep{2pt}
\begin{eqnarray} \label{ds2}
ds^2_{10}&=&R^2(-\cosh^2 \!\!\rho \;dt^2 + d \rho^2 +\sinh^2 \!\!\rho
\;d \Omega_3 )+\frac{R^2}{8}(d\tau +\frac{3}{2} \cos^2 \omega \,(d\psi
+\cos\theta \, d\phi))^2 +{}\nonumber\\ &&{}+\frac{3}{4}R^2(d\omega^2
+\sin^2\omega \cos^2 \omega
\,\sigma_3^2+\cos^2\omega\,(\sigma_1^2+\sigma_2^2))
\end{eqnarray}
where
\bd
\begin{array}{ll} 
\sigma_1+i\sigma_2=-\frac{i}{2}e^{i\psi}(d\theta-i\sin\theta \,d\phi) 
\\ \sigma_3=\frac{1}{2}(d\psi+\cos\theta\, d\phi) \end{array}
\ed while for the complex three--form we obtain
{\setlength\arraycolsep{2pt}
\begin{eqnarray} \label{G3}
G_3&=&\frac{3}{8\sqrt2}\, e^{i\alpha}e^{-i(\tau+\psi)}R^2 \cos\omega
\,
(d\theta+i\sin\theta\,d\phi)\wedge(d\omega+\frac{i}{2}\sin\omega\cos\omega
\,(d\psi+cos\theta \, d\phi))\wedge {}\nonumber\\ &&{}\wedge
(d\tau+\frac{3}{2}\cos^2\omega\, (d\psi+cos\theta \, d\phi))
\end{eqnarray}

The symmetry of this solution is $SU(3)\times U(1)$\footnote{The
$U(1)$ factor doesn't appear in the original work of Romans
\cite{Romans}. It was first argued to belong to the symmetry group of
this solution in \cite{Gunaydin}.}. The $SU(3)$ is a spatial symmetry,
it comes from the symmetry of the compact space $M_5$, while the
$U(1)$ is a subgroup of the $SU(1,1) \subset SL(2,\mathbb{R})$
symmetry of chiral supergravity \cite{Gunaydin} and can be associated
with translations of the $\alpha$ phase of the complex 3--form.

This solution has no supersymmetries, and is unstable. In fact there
is a scalar mode in the \textbf{6} of $SU(3)$ with $m^2=-\frac{40}{9}$
(in units of the $AdS$ radius) which is below the
Breitenlohner--Freedman bound \cite{BF}.

\vspace{12pt}

There is another way to build the solution $AdS_5\times M^5$. This
background is the ten--dimensional lifting of the solution
corresponding to one of the critical points of the scalar potential of
five--dimensional gauged supergravity \cite{KPW}. Each of these
critical points is characterized by the expectation value of a set of
scalar fields. After making the proper changes, we can use the same
formulas as \cite{PW1} for the embedding of the five--dimensional
solution into chiral ten--dimensional supergravity.

The general formula is
\begin{equation} \label{metPW}
ds^2_{10}=\Omega^2 ds^2_{AdS} + ds^2_5
\end{equation}
where 
\begin{gather}
\Omega^2=\xi \cosh\chi \nonumber\\
ds^2_5(\alpha,\chi)=\frac{a^2}{2}\frac{\sech\chi}{\xi}\lp dx^I
Q_{IJ}^{-1}dx^J\rp +\frac{a^2}{2}\frac{\sinh\chi\tanh\chi}{\xi^3}\lp
x^IJ_{IJ}dx^J\rp^2 \nonumber \\ Q_{IJ}={\mathrm {diag}}\lp
e^{-2\nu},e^{-2\nu},e^{-2\nu},e^{-2\nu}, e^{4\nu},e^{4\nu}\rp
\label{PWsol} \\ J_{IJ}=-J_{JI} \quad {\mathrm {and}} \quad
J_{14}=J_{23}=J_{65}=1\nonumber \\ \xi^2=x^IQ_{IJ}x^J \nonumber
\end{gather}
and $R_0$ is the radius of the round $S^5$ compactification, while
$\chi$ and $\nu$ are the scalar fields which determine the
five--dimensional solution. In our case they are
\begin{equation}
\label{critvev}\nu=0 \qquad {\mathrm {and}} \qquad
\chi=\frac{1}{2}\log\lp 2-\sqrt{3}\rp\end{equation} New complex
coordinates are defined based on the structure of {\it J}
\begin{equation} u^1=x^1+ix^4,\qquad u^2=x^2+ix^3, \qquad u^3=x^5-ix^6
\end{equation} which transform in the {\bf 3} of $SU(3)$.  We
parametrize them as
\begin{align} \label{u}
u^1 &=\cos\omega\, e^{i(P+\frac{\psi+\phi}{2})}\cos\frac{\theta}{2}
\nonumber\\ u^2 &=\cos\omega\,
e^{i(P+\frac{\psi-\phi}{2})}\sin\frac{\theta}{2} \\ u^3 &=\sin\omega\,
e^{iP} \nonumber
\end{align}
so that if we define $z_1=\frac{u^1}{u^3}$ and $z_2=\frac{u^2}{u^3}$,
these coordinates give the right parametrization (\ref{parcp}) of \cp.
Now we only need to make two remarks in order to obtain our solution
(\ref{ds2}). First of all the radius $R_0$ of $AdS_5$ in the round
$S^5$ compactification is different from our radius {\it R}:
(\ref{metPW}), (\ref{PWsol}) and (\ref{critvev}) show that
\begin{equation} R_0^2\,\,\sech\chi=\frac{2}{3}R^2 \end{equation} Also
the coordinate {\it P} has to be rescaled in order to obtain the
metric of the fibration over \cp $\,$as we wrote it above. The right
substitution is $P=\frac{\tau}{3}$.

This way of obtaining the solution is very useful when considering the
symmetries of the theory. If we write the metric (\ref{PWsol}) and
complex three--form in terms of the {\it u} coordinates \begin{eqnarray}
ds^2_5&=&\frac{2}{3}\,R^2 \sum_{k=1}^3 \lp du^k
d\bar{u}^k+\frac{1}{8}\lp \bar{u}^k du^k - u^k d\bar{u}^k\rp ^2 \rp \\
\label{G3uuu} 
G_3&=& i\,\frac{9}{4\sqrt{2}}\,e^{i\alpha}\,
d\bar{u}^1\wedge d\bar{u}^2\wedge d\bar{u}^3 \label{G3u}
\end{eqnarray}
the solution is manifestly $SU(3)\times U(1)$ symmetric.


\newsection{The Penrose limit}
\subsection{Taking the limit} \label{subsect:limit}
We now consider the Penrose limit of (\ref{ds2}). We consider the null
geodesic $\rho=\omega=\theta=0$ and scale our coordinates as
\begin{equation} \rho=\frac{r}{R}\qquad
\omega=\frac{2}{\sqrt{3}}\frac{\eta}{R}\qquad
\theta=\frac{4}{\sqrt{3}}\frac{\chi}{R} \end{equation} We define
\begin{equation} 
\begin{split} &\beta = \frac{\psi+\phi}{2} \\ &x^+ =
\frac{1}{2}\left(t+\frac{1}{2\sqrt{2}}(\tau+3\beta)\right) \\ &x^- =
\frac{R^2}{2}\left(t-\frac{1}{2\sqrt{2}}(\tau+3\beta)\right)
\end{split} \end{equation} and expand the metric keeping only 
{\it O(1)} terms
\begin{eqnarray} ds^2= &-&
4dx^+\,dx^--r^2(dx^+)^2+dr^2+r^2\,d\Omega_3+d\eta^2+\eta^2\,
d\beta^2+d\chi^2+\chi^2\, d\phi^2 + \nonumber\\ &-&
2\sqrt{2}dx^+(\eta^2\,d\beta+\chi^2\,d\phi^2) \end{eqnarray} 
There is a way to
write this metric which is far more intuitive in view of the
quantization of the string theory on this background. If we change
coordinates \begin{equation} \begin{split} \varphi_1 &= \phi-\sqrt{2} \lp
x^+-\frac{x^-}{R^2} \rp \\ \varphi_2 &= \beta-\sqrt{2} \lp
x^+-\frac{x^-}{R^2}\rp \end{split} \end{equation} the metric reads 
\begin{equation} 
\label{metpp}
ds^2=-4dx^+\,dx^-+\sum_{i=1}^{8}(dx^i)^2-\left(\sum_{i=1}^{4}(x^i)^2+2
\sum_{i=5}^{8}(x^i)^2\right)(dx^+)^2
\end{equation} 
where $x_1,\ldots ,x_4$ are along the spatial coordinates of the
$AdS$ part of the pp--wave and we defined
$\chi\,e^{i\varphi_1}=z_1=x^5+i\,x^6$ and
$\eta\,e^{i\varphi_2}=z_2=x^7+i\,x^8$.

We take the same limit on the 3-- and 5--forms, and obtain
\begin{equation} \label{forms}
\begin{split} &G_3 =2\,e^{i\alpha}\,dx^+\wedge dz_1 \wedge dz_2  
\\ &F_5 =-\frac{1}{\sqrt{2}}\,dx^+ \wedge (dx^1\wedge dx^2 \wedge dx^3 
\wedge dx^4 +dx^5\wedge dx^6 \wedge dx^7 \wedge dx^8) \end{split}
\end{equation}

The metric and the forms we have obtained from the Penrose limit of
the $SU(3) \times U(1)$ solution are, as one would expect, a solution
of the equations of motion of supergravity. They indeed satisfy the
relation \cite{CVJ} \bd \mathrm{tr} A =-8f^2 -2 \, \vert b \vert ^2
\ed where $A_{ij}$ are the masses of the bosonic zero modes that can
be read off the coefficents of the $(dx^+)^2$ terms in the metric,
e.g. $A_{ij}=\mathrm{diag}(1,1,1,1,2,2,2,2)$, and {\it b} and {\it f}
are the coefficients of $G_3$ and $F_5$ respectively.

As usual, the solution preserves the 16 supercharges
$\Gamma^+\epsilon=0$ \cite{Blau}. In Appendix \ref{app:susy} we check
that there are no other supersymmetries.


\newsection{String spectrum} \label{sec:spectrum}
\subsection{Bosonic sector}
The string theory on a pp--wave background is exactly solvable even
when there are non trivial NS--NS and R--R fields. The bosonic
spectrum doesn't get contributions from the R--R fields \cite{MT}, but
feels only the graviton and the NS--NS 3--form field strength. The
model we consider and its quantization are similar to \cite{Warner},
\cite{CVJ} and \cite{RT}.

We introduce a mass parameter {\it m} by scaling $x^+$ and $x^-$ as
\begin{equation} 
\label{rescaling} x^+\rightarrow m x^+ \qquad x^-\rightarrow
\frac{x^-}{m} \end{equation} Both $x^+$ and $x^-$ have now the
dimension of a length. We also decompose the complex 3--form $G_3$ as
\bd G_3=H_3^{NS}+i F_3^{RR}\ed
The background is then given by
\begin{eqnarray}
\label{background} &ds^2& = -4dx^+dx^-+\sum_{i=1}^{8}\lp dx^i\rp ^2 -
m^2 \lp \sum_{i=1}^{4}\lp x^i\rp ^2 +2\sum_{i=5}^{8}\lp dx^i\rp ^2
\rp\lp dx^+ \rp ^2 \\ &H_3^{NS}&=2m\, dx^+ \wedge \lp \cos\alpha \lp
dx^5\wedge dx^7-dx^6\wedge dx^8\rp-\sin\alpha \lp dx^6\wedge
dx^7+dx^5\wedge dx^8\rp\rp \nonumber \\ &F_3^{RR}&=2m\, dx^+\wedge\lp
\cos\alpha\lp dx^6\wedge dx^7+dx^5\wedge dx^8\rp+\sin\alpha\lp
dx^5\wedge dx^7-dx^6\wedge dx^8\rp\rp \nonumber\\
&F_5&=-\frac{m}{\sqrt{2}}\, dx^+\wedge\lp dx^1 \wedge dx^2 \wedge dx^3
\wedge dx^4 +dx^5 \wedge dx^6 \wedge dx^7 \wedge dx^8\rp \nonumber
\end{eqnarray} and the action for the bosonic sector is \cite{RT}
\begin{equation} S_B=\frac{1}{2\pi\alppr}\int d\tilde{\tau}
\int_0^{2\pi} d\sigma \, \tilde{{\cal L}}_B \end{equation} where,
after fixing the gauge $x^+=2\alppr \p \tilde{\tau}$, the
``lagrangian'' is given by \cite{RT} \begin{equation}
\label{lgrbtilde} \tilde{{\cal
L}}_B=\frac{1}{4}\!\lp\tilde{\partial_0}
X^i\rp^2-m^2\!\lp\alppr\p\rp^2 c_i
\!\left.X^i\right.^2-\frac{1}{4}\lp\!\left.X^j\right.'-2\alppr\p
H_{+ij}X^i\rp \! \left.X^j\right.'- 
2\alppr\p\lp\tilde{\partial_0} X^-
\!-\!\left. X^-\right.'\rp \end{equation} and
$c_i=(1,1,1,1,2,2,2,2)$. $\tilde{{\cal L}}_B$ doesn't have the
dimensions of an energy. This is because $\tilde{\tau}$ is not a
time--sized variable. A natural choice for defining a new time
coordinate is suggested by the gauge--fixing condition \begin{equation} 
\label{xgaugefix} \tau=2\alppr \p \tilde{\tau}=x^+\qquad \Rightarrow
\qquad \tilde{\partial}_0=\frac{\partial}{\partial\tilde{\tau}}
=2\alppr\p\frac{\partial}{\partial\tau}=2\alppr\p\partial_0
\end{equation} We also require that the lagrangian and the
gauge--fixing condition are consistent, thus we rescale $\tilde{{\cal
L}}_B$ in such a way that its functional derivative with respect to
$\dot{X}^-$ gives $p_-=-2p^+$.  We thus get for the lagrangian and
action, respectively
\begin{eqnarray} \label{lgrb} {\cal L}_B &=&\frac{\tilde{{\cal
L}}_B}{2\alppr^2\p}=\nonumber\\&=&
\frac{\p}{2}\sum_{i=1}^{8}\left.\dot{X}^i\right.^2-\frac{m^2\p}{2}
\sum_{i=1}^{8}c_i\left. X^i\right.^2-\frac{1}{8(\alppr)^2\p}
\sum_{i,j=1}^{8}\lp\delta_{ij}{X^i}^\prime-2\alppr\p
H_{+ij}X^i\rp{X^j}^\prime+ \nonumber \\ &
&-2\p\dot{X}^-+\frac{1}{\alppr}{X^-}^\prime \end{eqnarray} and
\begin{equation} S_B=\frac{1}{2\pi}\int d\tau \int_0^{2\pi} d\sigma \,
{\cal L}_B \end{equation}

We may now proceed in the evaluation of the bosonic string spectrum:
if we explicitly substitute $H_{+ij}$ into (\ref{lgrb}) then
\begin{eqnarray} {\cal L}_B=& &
\!\!\!\!\!\!\frac{\p}{2}\sum_{i=1}^{8}\left.\dot{X}^i\right.^2-
\frac{m^2\p}{2}\sum_{i=1}^{8}c_i\left. X^i\right.^2-
\frac{1}{8(\alppr)^2\p}\sum_{i,j=1}^{8}\left. {X^i}^\prime\right.^2
+\frac{m}{2\alppr}\lp \cos \alpha\lp X^5 {X^7}^\prime
+\right.\right.\nonumber\\ &-& \left.\left.X^7 {X^5}^\prime- X^6
{X^8}^\prime + X^8 {X^6}^\prime\rp -\sin \alpha \lp X^6 {X^7}^\prime -
X^7 {X^6}^\prime + X^5 {X^8}^\prime - X^8 {X^5}^\prime
\rp\rp+\nonumber \\ &-& 2\p\dot{X}^-+\frac{1}{\alppr}{X^-}^\prime
\end{eqnarray} Four bosonic modes are independent harmonic
oscillators, while the other four are magnetically coupled through the
$H_3$ field. The term in the $X^-$ field only gives a constraint on
the physical states, since it is linear in the field.  We introduce
three parameters \begin{equation} k\equiv \frac{\p}{2} \qquad a\equiv
\frac{1}{8(\alppr)^2\p} \qquad g\equiv \frac{m}{2\alppr}
\end{equation} to avoid as much confusion as possible in the following
expressions. We study the equations of motion of the fields $\lp
X^5,X^6,X^7,X^8\rp $; the ones for the four independent fields $\lp
X^1,X^2,X^3,X^4\rp $ are obtained by substituting $2m^2 \rightarrow
m^2$ and $g\rightarrow 0$ in the expressions for the interacting
degrees of freedom.

The Eulero--Lagrange equations read \begin{eqnarray}
&-&\ddot{X^5}=2m^2X^5-\frac{a}{k}{X^5}''-\frac{g}{k}\lp \cos\alpha
{X^7}'-\sin\alpha {X^8}'\rp \nonumber \\
&-&\ddot{X^6}=2m^2X^6-\frac{a}{k}{X^6}''+\frac{g}{k}\lp \cos\alpha
{X^8}'+\sin\alpha {X^7}'\rp \\
&-&\ddot{X^7}=2m^2X^7-\frac{a}{k}{X^7}''+\frac{g}{k}\lp \cos\alpha
{X^5}'-\sin\alpha {X^6}'\rp \nonumber \\
&-&\ddot{X^8}=2m^2X^8-\frac{a}{k}{X^8}''-\frac{g}{k}\lp \cos\alpha
{X^6}'+\sin\alpha {X^5}'\nonumber\rp \end{eqnarray} We build a
``vector'' $X(\sigma, \tau) \equiv \left\{
X^i\lp\sigma,\tau\rp\right\}$ with $i=5,6,7,8$ and expand it in
Fourier modes \begin{equation} X\lp\sigma+2\pi,\tau\rp =
X\lp\sigma,\tau\rp \qquad
X\lp\sigma,\tau\rp=\sum_{n=-\infty}^{+\infty}c_n(\tau)e^{in\sigma}
\end{equation} From the reality of $X$ it follows that
$\overline{c_{-n}}(\tau)=c_n(\tau)$. The system of equations of motion
becomes \begin{equation} \label{eqm} -\ddot{c}_n(\tau)=T_nc_n(\tau)
\end{equation} where \begin{equation} {\setlength\arraycolsep{6pt}
T_n=\lp\begin{array}{cccc} 2m^2+\frac{an^2}{k} & 0 &
-\frac{ign}{k}\cos\alpha & \frac{ign}{k}\sin\alpha \\ \\ 0 &
2m^2+\frac{an^2}{k} & \frac{ign}{k}\sin\alpha &
\frac{ign}{k}\cos\alpha \\ \\ \frac{ign}{k}\cos\alpha &
-\frac{ign}{k}\sin\alpha & 2m^2+\frac{an^2}{k} & 0 \\ \\
-\frac{ign}{k}\sin\alpha & -\frac{ign}{k}\cos\alpha & 0 &
2m^2+\frac{an^2}{k} \end{array}\rp} \end{equation} The eigenvalues of
this matrix are \begin{eqnarray} k_n^+ &=&
2m^2+\frac{an^2}{k}+\frac{g}{k}n \\ k_n^- &=&
2m^2+\frac{an^2}{k}-\frac{g}{k}n \end{eqnarray} each with multiplicity
2.  Since $T_n$ is a self--adjoint operator, $T_n^\dagger=T_n$, we
can choose a set of orthonormal eigenvectors of $T_n$ as a basis for
the four--dimensional space of the $c_n(\tau)$'s ($i=1,2$)
\begin{eqnarray} T_n e_n^{+(i)} &=& k_n^+ e_n^{+(i)} \nonumber \\ T_n
e_n^{-(i)} &=& k_n^- e_n^{-(i)} \\ \overline{e_n^{\pm (i)}} \cdot
e_n^{\pm (j)} &=& \delta^{+-} \delta^{ij}\nonumber \end{eqnarray} We
can write then \begin{equation} c_n(\tau)= B_n^{+(i)} e_n^{+(i)}+
B_n^{-(i)} e_n^{-(i)} \end{equation} where a sum over $i=1,2$ is
understood. Substituting into (\ref{eqm}) we find \begin{eqnarray}
-\ddot{B}_n^{+(i)}(\tau) &=& k_n^+ B_n^{+(i)}(\tau) \nonumber \\
-\ddot{B}_n^{-(i)}(\tau) &=& k_n^- B_n^{-(i)}(\tau) \end{eqnarray}
which have solution \begin{eqnarray} B_n^{+(i)}(\tau) &=& A_n^{+(i)}
e^{-i\omega_n^+ \tau} + D_n^{+(i)} e^{i\omega_n^+ \tau} \nonumber \\
B_n^{-(i)}(\tau) &=& A_n^{-(i)} e^{-i\omega_n^- \tau} + D_n^{-(i)}
e^{i\omega_n^- \tau} \end{eqnarray}
where \begin{equation} \omega_n^\pm=\sqrt{k_n^\pm} \end{equation} 
Imposing the condition on
the reality of {\it X}, we find a relation between the {\it A} and
{\it D} coefficients 
\begin{equation} \overline{A_{-n}^{\mp (i)}}=D_n^{\pm (i)} \end{equation}
We then have \begin{equation} \label{Xfield}
X(\sigma,\tau)=\sum_{n=-\infty}^{+\infty}\sum_{i=1}^{2}\lp\lp
A_n^{+(i)} e^{-i\omega_n^+ \tau} +
\overline{A_{-n}^{-(i)}}e^{i\omega_n^+ \tau}\rp\lp A_n^{-(i)}
e^{-i\omega_n^- \tau} + \overline{A_{-n}^{+(i)}}e^{i\omega_n^- \tau}
\rp\rp \end{equation} We define the momentum vector $\Pi (\sigma,\tau)=
\left\{\Pi_k\lp\sigma,\tau\rp \right\} = 2k \dot{X}(\sigma,\tau)$ with
$k=5,6,7,8$ and $\Pi_k(\sigma,\tau)\equiv\frac{\delta {\cal
L_B}}{\delta \dot{X}^k(\sigma,\tau)}$.  (\ref{Xfield}) and
$\Pi\lp\sigma,\tau\rp$ can be inverted to obtain the Fourier
coefficients $A_n^{\pm(i)}$ and $\overline{A_n^{\pm(i)}}$ \begin{eqnarray}
\label{fourcoeff}
A_n^{\pm(i)}&=&\frac{1}{2}\int_0^{2\pi}\frac{d\sigma}{2\pi}
\overline{e_n^{\pm (i)}} \cdot \lp X(\sigma,\tau)
+\frac{i}{2k\omega_n^\pm}\Pi(\sigma,\tau)\rp
e^{-in\sigma+i\omega_n^\pm\tau} \nonumber \\ \overline{
A_n^{\pm(i)}}&=&\frac{1}{2}\int_0^{2\pi}\frac{d\sigma}{2\pi} e_n^{\pm
(i)} \cdot \lp X(\sigma,\tau)
-\frac{i}{2k\omega_n^\pm}\Pi(\sigma,\tau)\rp
e^{in\sigma-i\omega_n^\pm\tau} \end{eqnarray} Quantization of the
bosonic fields is then achieved as usual by promoting $X(\sigma,\tau)$
and $\Pi(\sigma,\tau)$ to operators (the Fourier coefficients are also
promoted to operators $A_n^{\pm(i)}$ and
$\left. A_n^{\pm(i)}\right. ^\dagger$) and imposing canonical
commutation relations on them \begin{equation} \label{commrel} \left[
X^i(\sigma,\tau),\Pi_j(\sigma',\tau) \right] = i \,\delta
^i_{\phantom{i}j}\,\delta(\sigma-\sigma') \end{equation} By using
(\ref{fourcoeff}) and (\ref{commrel}) we can calculate the commutation
relations for the $A_n^{\pm(i)}$ and
$\left. A_n^{\pm(i)}\right. ^\dagger$ operators. Since we will
interpret them as creation and annihilation operators, we normalize
their commutators to one. We thus define ($\omega_n^\pm$ is always
strictly positive) \begin{eqnarray} a_n^{\pm(i)}&\equiv& \sqrt{8\pi k
\omega_n^\pm}\, A_n^{\pm(i)} \nonumber \\
\left. a_n^{\pm(i)}\right. ^\dagger &\equiv& \sqrt{8\pi k
\omega_n^\pm} \left. A_n^{\pm(i)}\right. ^\dagger \end{eqnarray} so
that \begin{equation} \left[
a_m^{+(i)},\left. a_n^{+(j)}\right.^\dagger \right] =
\delta_{mn}\,\delta^{ij} \qquad \mathrm{and} \qquad \left[
a_m^{-(i)},\left. a_n^{-(j)}\right.^\dagger \right] =
\delta_{mn}\,\delta^{ij} \end{equation} with all other commutators
equal to zero.

Substituting (\ref{Xfield}) and $\Pi\lp\sigma,\tau\rp$ into the
hamiltonian (${\cal L}_B^{5,6,7,8}$ stands for the part of the
lagrangian involving only the fields $X^5,\cdots,X^8)$
\begin{equation} H_B^{5,6,7,8}=\int_0^{2\pi}d\sigma\lp \Pi \cdot X -
{\cal L}_B^{5,6,7,8}\rp \end{equation} we find that \begin{equation}
H_B^{5,6,7,8}=\sum_{n=-\infty}^{+\infty}\lp\omega_n^+ \lp
\left.a_n^{+(1)}\right.^\dagger a_n^{+(1)}
+\left.a_n^{+(2)}\right.^\dagger a_n^{+(2)}\rp +\omega_n^- \lp
\left.a_n^{-(1)}\right.^\dagger a_n^{-(1)}
+\left.a_n^{-(2)}\right.^\dagger a_n^{-(2)}\rp\rp \end{equation} 
We
introduce the number operators
\begin{gather}
N_n^0\equiv \left.a_n^{0(1)}\right.^\dagger
a_n^{0(1)}+\left.a_n^{0(2)}\right.^\dagger
a_n^{0(2)}+\left.a_n^{0(3)}\right.^\dagger
a_n^{0(3)}+\left.a_n^{0(4)}\right.^\dagger a_n^{0(4)}\nonumber \\
N_n^+\equiv \left.a_n^{+(1)}\right.^\dagger
a_n^{+(1)}+\left.a_n^{+(2)}\right.^\dagger a_n^{+(2)} \qquad
N_n^-\equiv \left.a_n^{-(1)}\right.^\dagger
a_n^{-(1)}+\left.a_n^{-(2)}\right.^\dagger a_n^{-(2)}
\end{gather}
where the ``0'' quantities can be easily obtained from the $\pm$ ones
via the substitution mentioned above.  The complete bosonic
hamiltonian is given by \begin{equation} H_B=\sum_{n=-\infty}^{+\infty}\lp
\omega_n^0 N_n^0 +\omega_n^+N_n^+ +\omega_n^- N_n^- \rp \end{equation}


\subsection{Fermionic sector}

It comes out that in light--cone gauge, using the GS formalism, it is
possible to quantize the fermionic sector of superstring theory on a
pp--wave with non--zero 3--forms and 5--forms \cite{MT}, \cite{RT}. The
only contribution to the fermionic part of the lagrangian is given by
the supercovariant kinetic term for the two GS spinors \begin{equation} 
\label{lagrfertil} \tilde{{\cal
L}}_F=i\lp\eta^{ab}\delta_{IJ}-\epsilon^{ab}\rho_{3\,IJ}\rp
\tilde{\partial}_a x^m\bar{\theta}^I\Gamma_m(\hat{D}_b)^{JK}\theta^K
\end{equation} 
where \cite{MT} \cite{Warner} \begin{equation} \label{superder} \hat{D}_a
=\tilde{\partial}_a+\frac{1}{4}\tilde{\partial}_a
x^k\left[\lp\omega_{mnk}-\frac{1}{2}H_{mnk}\rho_3\rp\Gamma^{mn}-\lp
\frac{1}{12}F_{mnl}\Gamma^{mnl}\rho_1+ \frac{1}{120}F_{mnlpq}
\Gamma^{mnlpq} \rho_0\rp \Gamma_k\right]
\end{equation} 
\begin{equation} 
{\setlength\arraycolsep{4pt} \rho_0=\lp\begin{array}{cc} 0 & 1
\\ -1 & 0 \end{array} \rp\qquad\rho_1=\lp\begin{array}{cc} 0 & 1 \\ 1
& 0 \end{array} \rp\qquad \rho_3=\lp\begin{array}{cc} 1 & 0 \\ 0 & -1
\end{array} \rp } \end{equation} 
and $\theta^1$ and $\theta^2$ are 32--component
space--time Majorana spinors; since we are considering IIB
superstrings they have the same chirality. We choose
$\Gamma_{11}\theta^I=\theta^I$ to be consistent with the choices made
for the background gravitino and dilatino in Appendix
\ref{app:susy}\footnote{We explain our conventions on gamma--matrices
and spinors in Appendix \ref{app:conv}.}.

Fixing the light--cone gauge \begin{equation}
x^+=2\alppr\p\tilde{\tau}=\tau \qquad {\mathrm {and}} \qquad
\Gamma^+\theta^I=0 \end{equation} the supercovariant derivative can be
simplified to \begin{eqnarray}
\hat{D}_b^{\phantom{b}JM}\theta^M&=&\tilde{\partial}_b\theta^J-m
\alppr\p\delta_b^{\phantom{b}0}\left[\lp\cos\alpha\lp\Gamma^{57}-
\Gamma^{68}\rp-\sin\alpha\lp\Gamma^{67}
+\Gamma^{58}\rp\rp\rho_3^{\phantom{3}JM}\theta^M
+\phantom{\frac{1}{\sqrt{2}}}\right. \nonumber \\&{}& +
\left.\lp\cos\alpha\lp\Gamma^{67}+\Gamma^{58}\rp+\sin\alpha\lp
\Gamma^{57}-\Gamma^{68}\rp\rp\rho_1^{\phantom{1}JM}\theta^M \right. +
\\&{}& \left. -\frac{1}{\sqrt{2}}\lp\Gamma^{1234}+\Gamma^{5678}
\rp\rho_0^{\phantom{0}JM}\theta^M\right] \nonumber \end{eqnarray}

Using some gamma--matrices algebra and the chirality of the $\theta$
spinors it comes out that the only non--zero contribution to the sum
over {\it m} in (\ref{lagrfertil}) comes from the + term. As explained
in Appendix \ref{app:conv}, we use the light--cone gauge and the
chirality of $\theta^I$ to reduce the degrees of freedom of the
spinors, and write the lagrangian using 8--component spinors $S^I$ in
place of the 32--component ones $\theta^I$. Taking the same
normalization as for the bosonic lagrangian, we eventually get
\begin{equation}
\begin{split}
{\cal L}_F =\frac{2i}{\alppr} &
\left\{\phantom{\sqrt{2}}\!\!\!\!\!\!\!\!\!-2\alppr\p \lp
S^1\partial_0 S^1 + S^2\partial_0 S^2 \rp - S^1\partial_1 S^1 +
S^2\partial_1 S^2 +m\alppr\p\left[\phantom{\sqrt{2}}\!\!\!\!\!\!\!\!
S^1\lp\cos\alpha\lp\gamma^{57} +\right.\right.\right.\right.  \\ &\;\;
- \left.\left. \left.\left. \!\!\gamma^{68}\rp -
\sin\alpha\lp\gamma^{67}+\gamma^{58}\rp\rp S^1\! -S^2
\lp\cos\alpha\lp\gamma^{57}-\gamma^{68}\rp-\sin\alpha\lp\gamma^{67}+
\gamma^{58}\rp\rp
S^2 + \right.\right.  \\ &\;\;+ \left. \left.2
S^1\lp\cos\alpha\lp\gamma^{67}+\gamma^{58}\rp+\sin\alpha\lp\gamma^{57}-
\gamma^{68}\rp\rp
S^2+2\sqrt{2}\,S^2\,\gamma^{5678}\, 
S^1\right] \right\}
\end{split}
\end{equation}

We can now quantize the fermionic sector. First of all we write the
equations of motion
\begin{align} \label{eqnmotfer}
-i \dot{S}^1 &= \frac{i}{2\alppr\p}\partial_1 S^1 -\frac{im}{2}\left[
\lp
\cos\alpha\lp\gamma^{57}-\gamma^{68}\rp-\sin\alpha\lp\gamma^{67}+
\gamma^{58}\rp\rp
S^1 + \lp\cos\alpha\lp\gamma^{67}+\gamma^{58}\rp \nonumber
\right. \right.\\&\qquad\qquad\qquad \qquad
\left. \left. \sin\alpha\lp\gamma^{57}-\gamma^{68}\rp\rp S^2
-\sqrt{2}\gamma^{5678} S^2\right]\nonumber \\ -i
\dot{S}^2&=-\frac{i}{2\alppr\p}\partial_1 S^2 -\frac{im}{2}\left[
\lp\cos\alpha\lp\gamma^{57}-\gamma^{68}\rp-\sin\alpha\lp\gamma^{67}+
\gamma^{58}\rp\rp
S^2 +\lp\cos\alpha\lp\gamma^{67}+\right.\right.\right.\nonumber \\&
\qquad\qquad\qquad
\qquad+\left.\left.\left.\gamma^{58}\rp+\sin\alpha\lp\gamma^{57}-
\gamma^{68} \rp \rp
S^1 +\sqrt{2}\gamma^{5678} S^1\right]
\end{align}
and decompose the spinorial fields in Fourier modes
\begin{align}
&S^I(\sigma+2\pi,\tau) = S^I(\sigma,\tau) \\ &S^I(\sigma,\tau)=
\sum_{n=-\infty}^{+\infty}S_n^I(\tau)\,e^{in\sigma}
\end{align}

Writing the equations of motion for the normal modes we notice that
the $F_5$ interaction couples every component of one spinor to the
same component of the other spinor, without mixing different
components, while the 3--form interactions leave four modes unaffected
while coupling magnetically all the other components: $H_3$ acts
within the same spinor, while $F_3$ mixes components of the first
spinor with components of the second one, and viceversa. We then have
four components of each spinor ($S_{n(3)}^1$, $S_{n(4)}^1$,
$S_{n(5)}^1$, $S_{n(6)}^1$ and $S_{n(3)}^2$, $S_{n(4)}^2$,
$S_{n(5)}^2$, $S_{n(6)}^2$) which are coupled only through the 5--form
as \begin{equation} 
\begin{split}
-i
 \dot{S}_{n(k)}^1&=-\frac{n}{2\alppr\p}S_{n(k)}^1+
\frac{im}{\sqrt{2}}S_{n(k)}^2
 \\ -i \dot{S}_{n(k)}^2
 &=\frac{n}{2\alppr\p}S_{n(k)}^2-\frac{im}{\sqrt{2}}S_{n(k)}^1
\end{split}
\end{equation} 
Their frequencies can be found by diagonalizing the matrix
\begin{equation}
\label{smmat} \lp\begin{array}{cc} -\frac{n}{2\alppr\p} &
\frac{im}{\sqrt{2}} \\ -\frac{im}{\sqrt{2}} & \frac{n}{2\alppr\p}
\end{array}\rp \end{equation} 
which gives \begin{equation} \label{freq1} \omega_n^0 =
\pm\sqrt{\frac{m^2}{2}+\frac{n^2}{(2\alppr\p)^2}} \end{equation} The
other eight components feel both the complex 3--form and the 5--form
\begin{equation} -i\dot{\tilde{S}}_n=T_n\tilde{S}_n \end{equation}
where $\tilde{S_n} \equiv \lp S_{n(1)}^1 , S_{n(2)}^1 , S_{n(7)}^1 ,
S_{n(8)}^1, S_{n(1)}^2 , S_{n(2)}^2 , S_{n(7)}^2 , S_{n(8)}^2 \rp $
and \begin{equation} \label{Tn} T_n=
\begin{pmatrix} -\frac{n}{2\alppr\p} & 0 & -im\beta(\alpha) & 0 &
-\frac{im}{\sqrt{2}} & 0 & -m\beta(\alpha) & 0 \\ 0 &
-\frac{n}{2\alppr\p} & 0 & -im\beta(\alpha) & 0 & -\frac{im}{\sqrt{2}}
& 0 & -m\beta(\alpha)\\ im\bar{\beta}(\alpha) & 0 &
-\frac{n}{2\alppr\p} & 0 & -i\bar{\beta}(\alpha) & 0 &
-\frac{im}{\sqrt{2}} & 0 \\ 0 & im\bar{\beta}(\alpha) & 0 &
-\frac{n}{2\alppr\p} & 0 & -i\bar{\beta}(\alpha) & 0 &
-\frac{im}{\sqrt{2}} \\ \frac{im}{\sqrt{2}} & 0 & -m\beta(\alpha) & 0
& \frac{n}{2\alppr\p} & 0 & im\beta(\alpha) & 0 \\ 0 &
\frac{im}{\sqrt{2}} & 0 & -m\beta(\alpha) & 0 & \frac{n}{2\alppr\p} &
0 & im\beta(\alpha) \\ -m\bar{\beta}(\alpha) & 0 & \frac{im}{\sqrt{2}}
& 0 & -im\bar{\beta}(\alpha) & 0 & \frac{n}{2\alppr\p} & 0 \\ 0 &
-m\bar{\beta}(\alpha) & 0 & \frac{im}{\sqrt{2}} & 0 &
-im\bar{\beta}(\alpha) & 0 & \frac{n}{2\alppr\p}
\end{pmatrix}
\end{equation} and we have defined
$\beta(\alpha)\equiv\cos\alpha+i\sin\alpha$. The frequencies of the
normal modes can be obtained, again, by diagonalizing this matrix. The
result is \begin{eqnarray} \omega_n^- &=& \pm
\sqrt{\frac{n^2}{(2\alppr\p)^2}+
\frac{5}{2}m^2-2m\sqrt{\frac{n^2}{(2\alppr\p)^2}+m^2}}
\\ \omega_n^+ &=& \pm
\sqrt{\frac{n^2}{(2\alppr\p)^2}+
\frac{5}{2}m^2+2m\sqrt{\frac{n^2}{(2\alppr\p)^2}+m^2}}
\end{eqnarray}

To get the hamiltonian we organize the 16 eigenvalues in the following way
\begin{gather}
\omega_n^1=\omega_n^2=\omega_n^3=\omega_n^4=\omega_n^0 \nonumber\\
\omega_n^5=\omega_n^6=\omega_n^- \\
\omega_n^7=\omega_n^8=\omega_n^+\nonumber
\end{gather}
and the other eight ones are defined in the same way but with the
minus sign. The Majorana condition on the spinors $S^1$ and $S^2$
gives a relation between the positive-- and negative--frequency
components of the Fourier expansion $\Theta_n^j$. We obtain then
\begin{equation}
S(\sigma,\tau)=\sum_{n=-\infty}^{+\infty}\sum_{j=1}^{8}\lp
\Theta_n^j\, w_n^j\,e^{i\omega_n^j\tau}+\overline{\Theta}_{-n}^j\,
\overline{w}_{-n}^j\,e^{-i\omega_n^j \tau}\rp e^{in\sigma}
\end{equation} where $w_n^j$ are the eigenvalues of the kinetic matrix
built from the blocks (\ref{smmat}) and (\ref{Tn}).  We promote the
$\Theta_n^j$ and $\overline{\Theta}_n^j$ to operators $\Theta_n^j$ and
$\left.\Theta_n^j\right.^\dagger$ and define the new operators $b_n^j$
and $\left.b_n^j\right.^\dagger$ by requiring that imposing the
canonical anti--commutation relations for the spinor fields and
momentums \begin{equation} \left\{S^{I,i}(\sigma,\tau),\Pi^J_j(\sigma
',\tau)\right\}=i\delta^i_{\phantom{i}j}\delta(\sigma-\sigma
')\delta^{IJ} \qquad {\mathrm {for}}\quad i,j=1,\cdots,8
\end{equation} is equivalent to realizing a Clifford algebra with
$b_n^j$ and $\left.b_n^j\right.^\dagger$ \begin{equation}
\left\{b_m^i,\left.b_n^j\right.^\dagger\right\}=\delta_{mn}\delta^{ij}
\end{equation} The hamiltonian is given by \begin{equation} 
H_F=\sum_{n=-\infty}^{+\infty}\sum_{i=1}^8\omega_n^i
\left. S_n^i\right.^\dagger S_n^i \end{equation}


\subsection{Spectrum}

We summarize here the results we obtained in the two preceeding
subsections on the spectrum of superstring theory on this minimallly
supersymmetric pp--wave.  The total hamiltonian is given by 
\begin{equation} 
\begin{split}
H=H_B+H_F=\sum_{n=-\infty}^{+\infty}&\lp \omega_n^{B,0} N_n^{B,0}
+\omega_n^{B,+}N_n^{B,+} +\omega_n^{B,-}
N_n^{B,-}+\right.\\&\left.+\omega_n^{F,0} N_n^{F,0}
+\omega_n^{F,+}N_n^{F,+} +\omega_n^{F,-} N_n^{F,-}\rp \end{split} 
\end{equation} and 
\begin{equation} 
\label{bosmass} \begin{split} \omega_n^{B,0} &=
\sqrt{m^2+\frac{n^2}{\lp 2\alppr\p\rp ^2}} \\ \omega_n^{B,+} &=
\sqrt{2m^2+\frac{n^2}{\lp 2\alppr\p\rp ^2}+\frac{mn}{\alppr \p}} \\
\omega_n^{B,-} &= \sqrt{2m^2+\frac{n^2}{\lp 2\alppr\p\rp
^2}-\frac{mn}{\alppr \p}} \\ \omega_n^{F,0} &=
\sqrt{\frac{m^2}{2}+\frac{n^2}{(2\alppr\p)^2}} \\ \omega_n^{F,-} &=
\sqrt{\frac{n^2}{(2\alppr\p)^2}+
\frac{5}{2}m^2-2m\sqrt{\frac{n^2}{(2\alppr\p)^2}+m^2}}
\\ \omega_n^{F,+} &=
\sqrt{\frac{n^2}{(2\alppr\p)^2}+
\frac{5}{2}m^2+2m\sqrt{\frac{n^2}{(2\alppr\p)^2}+m^2}}
\end{split} \end{equation}

The physical states will also have to satisfy the constraint
\begin{equation}
\label{constraint} P=\sum_{n=-\infty}^{+\infty}nN_n=0 \end{equation} which
follows from the equations of motion of the $-$ coordinates.  We define
a vacuum $|0\rangle$ which is annihilated by all the bosonic and
fermionic destruction operators, and build string states by applying
the creation operators on it, taking care of satisfying the constraint
(\ref{constraint}).

As can be noted from the masses (\ref{bosmass}) of the bosonic modes,
the tachyon that made the ten--dimensional background unstable is not
present in the string spectrum after the Penrose limit. This could
have been guessed even before calculating the frequencies of the
bosonic modes, and is due to the fact that when we put the superstring
theory on the pp--wave limit of a supergravity compactification, we
are keeping only those states in string theory which have very
large momentum along the geodesic we take the Penrose limit on,
i.e. we are considering high--order Kaluza--Klein states. We argue
that the behaviour of the scalars of our theory is not much different
(at least qualitatively) from the $AdS_5\times S^5$ solution, where
the KK--angular momentum contribution to the mass of the scalars in
any representation of $SU(4)$ is eventually dominant over the other
contributions \cite{mass}. Thus, in the limit of large {\it J}, the KK
contribution to the energy of the states will drive the mass of the
tachyon to a positive value. This is analogous to what happens in the
Penrose limit of type IIB string theory on $AdS_5\times T^{p,q}$
\cite{Klebanov} and of type 0 string theory on $AdS_5\times S^5$
\cite{BCGZ}.


\subsection{Zero--point energy}

Despite the solution has 16 supersymmetries, the masses of the bosons
and fermions are not equal, and the string theory will have a
non--vanishing zero--point energy given by \begin{equation} \label{E0}
E_0=\sum_{n=-\infty}^{+\infty}\left. 
E_0\right._n=\sum_{n=-\infty}^{+\infty}\lp
2\omega_n^{B,0}+\omega_n^{B,-}+\omega_n^{B,+}-2\omega_n^{F,0}
-\omega_n^{F,-}-\omega_n^{F,+}\rp
\end{equation} 
Other examples of the same phenomenon include \cite{Klebanov},
\cite{BCGZ} and \cite{FT}.  The series (\ref{E0}) is convergent and we
can approximate it with the integral \begin{equation} \begin{split}
\int_{-\infty}^{+\infty}dx &\lp 2\sqrt{m^2+\frac{x^2}{\lp 2\alppr\p\rp
^2}} -2\sqrt{\frac{m^2}{2}+ \frac{x^2}{(2\alppr\p)^2}} +
\sqrt{2m^2+\frac{x^2}{\lp 2\alppr\p\rp ^2}+\frac{mx}{\alppr \p}}+
\right.\\ &\left. +\sqrt{2m^2+\frac{x^2}{\lp 2\alppr\p\rp
^2}-\frac{mx}{\alppr \p}} - \sqrt{\frac{x^2}{(2\alppr\p)^2}+
\frac{5}{2}m^2-2m\sqrt{\frac{x^2}{(2\alppr\p)^2}+m^2}}\right. \\
&\left.-\sqrt{\frac{x^2}{(2\alppr\p)^2}+
\frac{5}{2}m^2+2m\sqrt{\frac{x^2}{(2\alppr\p)^2}+m^2}}\,\rp
\end{split} \end{equation}
where we have substituted the discrete variable {\it n} with a
continuos one {\it x}. This integral can be evaluated numerically, and
gives a positive result \begin{equation} E_0\sim 2m^2\alppr\p
\end{equation}

It is interesting to compare this with another non--supersymmetric
example \cite{BCGZ}. There the zero--point energy $E_0/m$ vanishes for
$m\alppr\p\rightarrow +\infty$ (perturbative limit in the dual gauge
theory), while is unbounded from below in the limit
$m\alppr\p\rightarrow 0$ (supergravity limit). In this case we have
the opposite behaviour, $E_0/m$ goes to 0 in the supergravity limit and
diverges in the perturbative limit.


\newsection{Dual gauge theory}

The gauge theory dual to the compactification of \cite{Romans} is
${\cal N}=4$ $SU(4)_R$ super Yang-Mills deformed by a mass term for
one of the four fermions in the adjoint of the gauge group $SU(N)$
\cite{GPPZ} \cite{DZ}. The compactification we are considering is
indeed obtained from $AdS_5\times S^5$ by turning on a complex
3--form. Complex 3--forms are in the {\bf 10} of $SU(4)$ and couple to
boundary operators which are bilinears in the four fermions that
belong to the spectrum of ${\cal N}=4$ SYM. In particular the 3--form
(\ref{G3uuu}) we turned on is in the singlet representation of
$SU(3)\subset SU(4)$ and thus can only couple to a mass term for one
of the fermions. We choose $\lambda_4$ as the fermion that gets a
mass. Thus the gauge theory lagrangian is given by \begin{equation}
{\cal L}={\cal L_N}_{=4}+m\,\mathrm{Tr}\lp\lambda_4\lambda_4+
\bar{\lambda}_4\bar{\lambda}_4\rp
\end{equation}

The fermion mass term is a relevant operator which drives a RG flow
from the ${\cal N}=4$ UV fixed point to an IR fixed point where all
supersymmetries are broken. Nothing will prevent scalar fields to gain
mass through radiative corrections and the effective infrared theory
will be made up only of three massless fermions and $SU(N)$ gauge
fields, all in the adjoint of the gauge group. The authors of
\cite{Distler} argued that, as a consequence of the instability of
this fixed point, the chiral symmetry $SU(3)$ is dynamically broken
down to $SO(3)$. Since our model is stable after the Penrose limit, we
believe that the subset of operators we are considering and their
symmetries are well described by the pp--wave background.

Let us consider the symmetries preserved by the limit. In section
\ref{subsect:limit} we redefined the angle coordinates of the original
space
\begin{equation} 
\begin{split}
t &= m x^+ +\frac{x^-}{mR^2} \\ \phi &= \sqrt{2}\, \lp m
x^+-\frac{x^-}{mR^2}\rp +\varphi_1 \\ \psi &= \sqrt{2}\, \lp m
x^+-\frac{x^-}{mR^2}\rp +2\varphi_2-\varphi_1 \\ \tau &=-\sqrt{2}\,
\lp m x^+-\frac{x^-}{mR^2}\rp -3\varphi_2 \end{split} \end{equation}
thus the hamiltonian and the light--cone momentum are given by
\begin{equation} \label{der}
\begin{split} H&=-\frac{p_+}{m}=\frac{i}{m}\frac{\partial}{\partial
x^+}=i\frac{\partial}{\partial t}- i\sqrt{2}
\left(\frac{\partial}{\partial \tau}-\frac{\partial}{\partial
\psi}-\frac{\partial}{\partial \phi} \right)=\Delta-\sqrt{2}\lp
J_{\tau-\psi}-J_{\phi}\rp \\ 2p^+&=-mp_-=im \frac{\partial}{\partial
x^-}=\frac{1}{R^2}\lp i\frac{\partial}{\partial t}+ i\sqrt{2}
\left(\frac{\partial}{\partial \tau}-\frac{\partial}{\partial
\psi}-\frac{\partial}{\partial \phi}
\right)\rp=\frac{\Delta+\sqrt{2}\lp
J_{\tau-\psi}-J_{\phi}\rp}{R^2}\end{split} \end{equation} where $J_{\tau -
\psi}\equiv
i\lp\frac{\partial}{\partial\tau}-\frac{\partial}{\partial\psi}\rp$
and $J_{\phi}\equiv i\frac{\partial}{\partial\phi}$. As manifest from
(\ref{u}) $J_{\tau-\psi}-J_{\phi}$ is a $U(1)$ generator of $SU(3)$.

The boundary operator $\mathrm{Tr}\,\lambda_a\lambda_b+ \mathrm{h.c.}$
under $SU(4)\rightarrow SU(3)$ decomposes as \begin{equation}
\mathbf{10}\quad\rightarrow\quad\mathbf{1+3+6} \end{equation} As we
have already said, the singlet operator
$\mathrm{Tr}\,\lambda_4\lambda_4$ couples to\footnote{The coefficients
in front of the 3--forms are not relevant in the calculation of the
charges of the fermions, and we will ignore them. The couplings will
be valid up to a constant coefficient.}  $d\bar{u}_1\wedge
d\bar{u}_2\wedge d\bar{u}_3$, while $\mathrm{Tr}\,\lambda_i\lambda_j$
to $du_i\wedge d\bar{u}_{j} \wedge d\bar{u}_{k}$. From this we obtain
the charges of Table \ref{table:charges}, where $J\equiv
J_{\tau-\psi}-J_{\phi}$.
\begin{table}[t] 
\begin{center}
\begin{tabular}{|c|r @{} l|r @{} l|r @{} l|}
\hline & \multicolumn{2}{c|}{$J_{\tau -\psi}$} &
\multicolumn{2}{c|}{$J_{\phi}$} & \multicolumn{2}{c|}{$J$} \\ \hline
$\lambda_1$ &$-$& $\frac{1}{6}$ & & $\frac{1}{2}$ &$-$&
$\frac{2}{3}$\\ \hline $\bar{\lambda}_1$ & & $\frac{1}{6}$ &$-$&
$\frac{1}{2}$ & & $\frac{2}{3}$\\ \hline $\lambda_2$ &$-$&
$\frac{1}{6}$ &$-$& $\frac{1}{2}$ & & $\frac{1}{3}$\\ \hline
$\bar{\lambda}_2$ & & $\frac{1}{6}$ & & $\frac{1}{2}$ &$-$&
$\frac{1}{3}$\\ \hline $\lambda_3$ & & $\frac{1}{3}$ & & $0$ & &
$\frac{1}{3}$\\ \hline $\bar{\lambda}_3$ &$-$ & $\frac{1}{3}$ & & $0$
&$-$& $\frac{1}{3}$\\ \hline
\end{tabular} \caption{The charges of the fermionic massless fields 
which make up the spectrum of the 
effective theory at the IR fixed 
point.\label{table:charges}}\end{center} \end{table}
By looking at the charges it is evident that at the $n=0$ level the
symmetry conserved by the Penrose limit is $SU(2)\times U(1)\subset
SU(3)$, where $SU(2)$ rotates the fermions
$\lp\lambda_2,\,\lambda_3\rp$.

Since there are no supersymmetries, and the supergravity approximation
gives results only on some low--lying states, we can only make
conjectures on the spectrum of the fields, which we leave for future
work to verify. The $SU(3)$ symmetry of the original theory ensures
that the fields $\lambda_1,\lambda_2,\lambda_3$ and, separately,
$\bar{\lambda}_1,\bar{\lambda}_2,\bar{\lambda}_3$ will have the same
dimension. Thus looking at Table \ref{table:charges} it seems a good
guess to choose as the building block of the vacuum the field
$\bar{\lambda}_1$. Its charge is the largest among all the fields,
which means that if we take as a first approximation to the energy of
the operator the sum of its $\Delta$ and $J$,
Tr$\bar{\lambda}_1^{\phantom{1}2J}$ will be the one with the smallest
energy. Moreover $\bar{\lambda}_1$ is a singlet of the $SU(2)$
symmetry group of the free theory, feature we would expect from the
vacuum. Thus we take \begin{equation} \label{eq:vac} \vac=\mathrm{Tr}\,\lp
\bar{\lambda}_1^{\phantom{1}2J}\rp \end{equation}

The four bosonic states with $H=1$ are obtained, as usual in this kind
of theories, by applying the gauge--covariant derivative along one of
the four space--time directions to a scalar pair
$\lp\bar{\lambda}_1\bar{\lambda}_1\rp$. The covariant derivative adds
a unit to the dimension, while leaving the charge unchanged.
\begin{equation} H=1 \quad \rightarrow \quad
\mathrm{Tr}\lp\bar{\lambda}_1^{\phantom{1}2J-2}D_i\lp
\bar{\lambda}_1\bar{\lambda}_1\rp\rp \end{equation}

We argue the four bosonic states with $H=\sqrt{2}$ are obtained by
substituting one of the scalars $\lp\bar{\lambda}_1\bar{\lambda}_1\rp$
with one of the Goldstone bosons of the symmetries that were broken in
$SU(3)\rightarrow SU(2)\times U(1)$. The four generators of these
broken symmetries give rise to the operators
\begin{equation} \label{Goldstone}\begin{split}
&\mathrm{Tr}
\lp\bar{\lambda}_1^{\phantom{1}2J-2}\lp\bar{\lambda}_1\bar{\lambda}_2+
\bar{\lambda}_2\bar{\lambda}_1\rp\rp
\qquad
\;\,\mathrm{Tr}\lp\bar{\lambda}_1^{\phantom{1}2J-2}\lp\bar{\lambda}_1
\bar{\lambda}_3+\bar{\lambda}_3\bar{\lambda}_1\rp\rp\\&\mathrm{Tr}\lp
i\bar{\lambda}_1^{\phantom{1}2J-2}\lp\bar{\lambda}_1\bar{\lambda}_2-
\bar{\lambda}_2\bar{\lambda}_1\rp\rp
\qquad\mathrm{Tr}\lp
i\bar{\lambda}_1^{\phantom{1}2J-2}\lp\bar{\lambda}_1\bar{\lambda}_3-
\bar{\lambda}_3\bar{\lambda}_1\rp\rp\end{split}
\end{equation} This identification is strengthened by the following
consideration: if we substitute a $\bar{\lambda}_1$ field with a
$\bar{\lambda}_2$ or $\bar{\lambda}_3$ fermion, we are subtracting a
$2/3$ charge and adding a $-1/3$ one. Since these three fermions have
the same $\Delta$, if we naively add up the charges and dimension of
the new state we find that, because of (\ref{der}), $H=\sqrt{2}$.

The fermionic states should be built by adding a fermion to one of the
states. The six states with $H=1/\sqrt{2}$ could be realized by adding
to (\ref{eq:vac}) one of the fermions
$\bar{\lambda}_1,\bar{\lambda}_2,\bar{\lambda}_3$. Each of them has
two degrees of freedom, since they are Weyl fermions, giving a total
of six states. If this guess will prove to be right, the other two
states with $H=3/\sqrt{2}$ are to be expected to come from the
insertion of a $\bar{\lambda_4}$ (or the corresponding combination of
fields in the effective theory). When we integrate it out in the IR
fixed point, this field is presumably substituted by a trilinear in
the other three fermions (it is the easiest way to build a state with
$J=0$), giving an idea of why the energy of the two states
$\mathrm{Tr}\,\lp \bar{\lambda}_1^{\phantom{1}2J} \bar{\lambda_4}\rp$
should be three times the energy of the states $\mathrm{Tr}\,\lp
\bar{\lambda}_1^{\phantom{1}2J} \bar{\lambda_k}\rp$.

We define $\lambda_{\sss eff}=\frac{\lambda}{J^2}$. The stringy
operators are probably obtained from the $n=0$ ones by adding phases
as in \cite{BMN}. The dimensions have a common zero--point
contribution \begin{equation} \lp\Delta-J\rp_0=\frac{E_0}{m}\sim
\frac{1}{\sqrt{\lambda_{\sss eff}}}\end{equation} From (\ref{bosmass}),
(\ref{der}) and $R^4=4\pi g N\alppr^2$, 
the perturbative expansion for
the dimension of the single impurity operators reads 
\begin{equation} 
\begin{split}
& \lp\Delta-J\rp
_n^{B,0}=\sqrt{1+\frac{\pi}{2}\frac{n^2\lambda}{J^2}}=1+
\frac{\pi}{4}\,n^2\,\lambda_{\sss
eff}+O(\lambda_{\sss eff}^2) \\ \vspace{10pt} & \lp\Delta-J\rp
_n^{B,-}\!=\!\sqrt{2}\,
\sqrt{1-\sqrt{\frac{\pi}{2}}\frac{n\sqrt{\lambda}}{J}+
\frac{\pi}{4}\frac{n^2\lambda}{J^2}}=\sqrt{2}\lp
1\!-n{\textstyle \sqrt{\frac{\pi}{8}}}\sqrt{\lambda_{\sss
eff}}+\!\frac{\pi}{16}n^2\lambda_{\sss eff}+O(\lambda_{\sss
eff}^{3/2})\rp \\ \vspace{10pt} & \lp\Delta-J\rp
_n^{B,+}\!=\!\sqrt{2}
\,\sqrt{1+\sqrt{\frac{\pi}{2}}\frac{n\sqrt{\lambda}}{J}+
\frac{\pi}{4}\frac{n^2\lambda}{J^2}}=\sqrt{2}\lp
1\!+n{\textstyle \sqrt{\frac{\pi}{8}}}\sqrt{\lambda_{\sss
eff}}+\!\frac{\pi}{16}n^2\lambda_{\sss eff}+O(\lambda_{\sss
eff}^{3/2})\rp \\ \vspace{10pt} & \lp\Delta-J\rp
_n^{F,0}=\frac{1}{\sqrt{2}}\sqrt{1+\pi\frac{n^2\lambda}{J^2}}=
\frac{1}{\sqrt{2}}\lp
1 +\frac{\pi}{2}\,n^2\,\lambda_{\sss eff}+ O(\lambda_{\sss eff}^2)\rp
\\ \vspace{10pt} & \lp\Delta-J\rp
_n^{F,-}=\sqrt{\frac{5}{2}-2\sqrt{1+\frac{\pi}{2}\frac{n^2\lambda}{J^2}}
+\frac{\pi}{2}\frac{n^2\lambda}{J^2}}
=\frac{1}{\sqrt{2}}\lp 1+\frac{\pi^2}{16}\,n^4\,\lambda_{\sss eff}^2
+O(\lambda_{\sss eff}^3)\rp \\ \vspace{10pt} & \lp\Delta-J\rp
_n^{F,+}=\sqrt{\frac{5}{2}+2\sqrt{1+\frac{\pi}{2}\frac{n^2\lambda}{J^2}}
+\frac{\pi}{2}\frac{n^2\lambda}{J^2}}=\frac{3}{\sqrt{2}}\lp
1+\frac{\pi}{9}\,n^2\,\lambda_{\sss eff} + 
O(\lambda_{\sss eff}^2) \rp
\vspace{10pt}
\end{split}
\end{equation}

Some comments are in order. First of all we notice that the
zero--point contribution to $\Delta-J$ is divergent in the
perturbative limit. $\lp\Delta-J\rp_0$ is a constant common to all
operators, and perturbation theory should still allow us to calculate
the difference between the dimension of an operator and that of the
vacuum. We also find that in the expansions for the $H=\sqrt{2}$
scalars the first contribution is of order $\sqrt{\lambda_{\sss
eff}}$. This seems to be an original feature of our model, and from
\cite{Gross} we believe it suggests that the quantity
$e^{\frac{2\pi}{J}}-e^{-\frac{2\pi}{J}}$ should appear in the leading
coefficient of the perturbation expansion, indicating that moving an
impurity in one direction or the other, in Feynman graphs, should give
different contributions. It must also be noticed that the dimension of
two fermionic modes doesn't get corrections from one--loop graphs, and
its expansion starts at the second order. This same feature was found
in \cite{CVJ}.

We leave for future work a firmer analysis of the gauge theory.

\vspace{12pt}
\noindent {\bf Acknowledgements}

\vspace{6pt}
\noindent We would like to thank Luciano Girardello and Alberto
Zaffaroni for suggesting the problem under study and for useful
discussions. This work was partially supported by INFN and MURST under
contract 2001-025492, and by the European Commission TMR program
HPRN-CT-2000-00131, in association to the University of Padova.


\setcounter{section}{0} \setcounter{subsection}{0}

\appendix
\setcounter{equation}{0}
\renewcommand{\theequation}{\Alph{section}.\arabic{equation}}

\newsection{Appendix: $\Gamma$ matrices and conventions}
\label{app:conv} The metric signature is $\eta^{MN}=(-,+,+,\ldots
,+)$. When we write antisymmetric forms in component notation we use
the following normalization \bd
\omega_p=\frac{1}{p!}\,\omega_{1,\ldots, p}\,dx^1\wedge \ldots \wedge
dx^p \ed

We adopt the same conventions as \cite{Metsaev} for $\Gamma$--matrices
and indicate with $\Gamma^M$ the $32\times 32$--component
gamma--matrices, and with $\tilde{\gamma}^M$ the $16\times
16$--component ones.  \begin{equation} \Gamma^M=\lp\begin{array}{cc} 0
& \tilde{\gamma}^M \\ \bar{\tilde{\gamma}}^M & 0 \end{array} \rp
\end{equation} \begin{equation} \left\{\Gamma^M,\Gamma^N\right\}=
2\eta^{MN}\qquad\qquad\tilde{\gamma}^M\bar{\tilde{\gamma}}^N+
\tilde{\gamma}^N\bar{\tilde{\gamma}}^M=2\eta^{MN}
\end{equation} \begin{equation} \label{gamma16} \tilde{\gamma}^M=\lp
1,\tilde{\gamma}^i,\tilde{\gamma}^9\rp\qquad\bar{\tilde{\gamma}}^M=\lp
-1,\tilde{\gamma}^i,\tilde{\gamma}^9\rp \end{equation} where
$M,N=0,1,2,\ldots ,9$ and $i,j=1,2,3, \ldots ,8$.

We adopt the Majorana representation, $C=\Gamma^0$, so we can choose
all $\tilde{\gamma}^M$ to be real and symmetric, and assume the
normalization \begin{equation} \label{gamma11}
\Gamma_{11}\equiv\Gamma^0\Gamma^1\ldots \Gamma^9=\lp\begin{array}{cc}
1 & 0 \\0 & -1\end{array}\rp \end{equation}

We define $\Gamma^{M_1 \ldots M_p}$ as the antisymmetrized product of
$\Gamma$ matrices with the same normalization as forms \bd
\Gamma^{M_1 \ldots M_p}=\frac{1}{p!}\,\varepsilon_{M_1M_2 \ldots
M_p}\,\Gamma^{M_1}\Gamma^{M_2}\ldots
\Gamma^{M_p}=\Gamma^{M_1}\Gamma^{M_2}\ldots \Gamma^{M_p} \ed where the
last equation is valid if and only if all the indices are different
(otherwise the matrix equals zero).

\vspace{12pt}
\noindent {\it 32-- $\rightarrow$ 8--component spinor decomposition}
\vspace{12pt}

Because of the normalization we chose for $\Gamma_{11}$ and the
condition on the space--time GS spinors:
$\Gamma_{11}\theta^I=\theta^I$, the 32--component spinor has only 16
non--zero components \begin{equation}
\theta^I=\lp\begin{array}{c}\tilde{\theta}^I \\ 0 \end{array}\rp
\end{equation} where $\tilde{\theta}^I$ is a 16--component Majorana
spinor. Moreover, not all of the components of $\tilde{\theta}^I$ are
physical degrees of freedom, since we still have to take into account
the light--cone gauge. We define \begin{equation} \Gamma^+\equiv
\frac{\Gamma^0+\Gamma^9}{2}=\lp\begin{array}{cc} 0 & \tilde{\gamma}^+
\\ \bar{\tilde{\gamma}}^+ & 0 \end{array}\rp \qquad {\mathrm {and}}
\qquad\Gamma^-\equiv \frac{\Gamma^0-\Gamma^9}{2}=\lp\begin{array}{cc}
0 & \tilde{\gamma}^- \\ \bar{\tilde{\gamma}}^- & 0 \end{array}\rp
\end{equation} but now
\begin{equation} 
\Gamma_{11}\theta=\theta \qquad \Rightarrow \qquad \Gamma^0\theta= \lp
\begin{array}{c} 0 \\ -\tilde{\theta} \end{array}\rp \quad {\mathrm
{and}} \quad \Gamma^9\theta= \lp \begin{array}{c} 0 \\
\tilde{\gamma}^9\tilde{\theta} \end{array}\rp \end{equation} thus
\begin{equation}
\begin{split}
\Gamma^+\theta&=0 \qquad \Leftrightarrow \qquad
\bar{\tilde{\gamma}}^+\tilde{\theta}=0 \qquad \Leftrightarrow \qquad
\tilde{\gamma}^9\tilde{\theta}=\tilde{\theta} \\ \Gamma^-\theta&=0
\qquad \Leftrightarrow \qquad \bar{\tilde{\gamma}}^-\tilde{\theta}=0
\qquad \Leftrightarrow \qquad
\tilde{\gamma}^9\tilde{\theta}=-\tilde{\theta}
\end{split}
\end{equation} If we decompose the 16--component spinors as
$\tilde{\theta}=\lp\begin{array}{c} \chi^+ \\ \chi^- \end{array} \rp$
then \begin{equation} \chi^+=0 \quad \Leftrightarrow
\quad\bar{\tilde{\gamma}}^-\tilde{\theta}=0 \qquad {\mathrm {and}}
\qquad \chi^-=0 \quad \Leftrightarrow
\quad\bar{\tilde{\gamma}}^+\tilde{\theta}=0 \end{equation} and we can
represent $\tilde{\gamma}^9$ as $\lp \begin{array}{cc}1_8 & 0 \\ 0 &
-1_8
\end{array} \rp$.

The light--cone and chirality conditions are thus equivalent in our
representation to imposing that only the first 8 components (which
constitute a spinor {\it S} on their own) of a 32--component spinor
are non--zero.  A representation of the $SO(8)$ algebra of
gamma--matrices $\gamma^i$ can be built on these spinors (we will do
this in the next subsection), and in particular it is found that for
$i_1,\cdots,i_p=1,\cdots,8$
\begin{gather}
\Gamma^{i_1\cdots i_p}\theta \; \rightarrow \;\gamma^{i_1\cdots i_p}S
\\ \Gamma^0\theta \; \rightarrow \; -S \qquad \Gamma^9\theta \;
\rightarrow \; S \end{gather} In particular \begin{equation}
\gamma^{12\cdots 8}S=S
\end{equation} but since $\lp\gamma^{1234}\rp^2=1$ we have 
\begin{equation}
\gamma^{1234}S=\gamma^{5678}S \end{equation}

\vspace{12pt}
\noindent {\it SO(8) gamma--matrices algebra representation on
8--component spinors}
\vspace{12pt}

To build a realization of the {\it SO(8)} gamma--matrices algebra on
the 8--component spinors we define the combinations ($j=1,2,3,4$)
\begin{equation}
\begin{split}
a_j&=\frac{1}{2}\lp\gamma^{2j-1}-i\gamma^{2j}\rp \\
a_j^\dagger&=\frac{1}{2}\lp\gamma^{2j-1}+i\gamma^{2j}\rp\end{split}
\end{equation}
Because of the algebra $\{\gamma^i,\gamma^j\}=2\delta^{ij}$ the
operators which we just defined realize a Clifford algebra
\begin{equation} \left\{ a_j^\dagger,a_k\right\}=\delta_{jk} \qquad
\left\{ a_j^\dagger,a_k^\dagger\right\}=0 \qquad \left\{
a_j,a_k\right\}=0 \end{equation} and in particular $a_j^\dagger
a_j^\dagger=a_ja_j=0$.  We can interpret the $a_j$ and $a_j^\dagger$
as creation and annihilation operators on an 8--dimensional vector
space. We define the vacuum $|0\rangle$ as the state which is
annihilated by all the $a_j$ operators and build vectors by applying
the $a_j^\dagger$ operators on it and represent spinors as
\begin{equation} S=\lp\begin{array}{c} S_1 \\ S_2 \\ S_3 \\ S_4\\S_5
\\ S_6 \\ S_7 \\ S_8\end{array}\rp=\lp\begin{array}{c}\vac \\
a_1^\dagger a_2^\dagger\vac \\a_1^\dagger a_3^\dagger\vac \\
a_1^\dagger a_4^\dagger\vac \\a_2^\dagger a_3^\dagger\vac \\
a_2^\dagger a_4^\dagger\vac \\a_3^\dagger a_4^\dagger\vac \\
a_1^\dagger a_2^\dagger a_3^\dagger a_4^\dagger
\vac\end{array}\rp\end{equation}


\newsection{Supersymmetries} \label{app:susy} Following closely
\cite{Blau} and \cite{CVJ} we take the variation equations for the
dilatino and gravitino from \cite{susy} and set them to zero:
\begin{equation}
\label{susyeqns} \begin{split} &\delta \lambda = -\frac{i}{4} \Gs_{3}
\epsilon \\ &\delta \psi_M = D_M \epsilon + \frac{i}{4} \Fs_{5}
\Gamma_M \epsilon - \frac{1}{16} (2 \Gs_{3} \Gamma_M +\Gamma_M
\Gs_{3}) \epsilon^* \end{split} \end{equation} where $D_M
\epsilon=\partial_M \epsilon
+\frac{1}{4}\omega_M^{\phantom{M}NP}\Gamma_{NP}\,\epsilon$.  We take
$\Gamma_{11}\psi_M=\psi_M$ and $\Gamma_{11}\lambda=-\lambda$, so that
$\Gamma_{11}\epsilon=\epsilon$.  By choosing the vielbeins
\begin{equation} e^i=dx^i \qquad e^+=dx^+ \qquad e^-=dx^-+\frac{1}{4}
\left(\sum_{i=1}^{4}(x^i)^2+2\sum_{i=5}^{8}(x^i)^2\right)\,dx^+
\end{equation} the metric (\ref{metpp}) can be written as
\begin{equation} ds^2=-4e^+e^-+\sum_{i=1}^{8}(e^i)^2 \end{equation} We
find that the only non--zero components of the spin connection
$\omega_M^{\phantom{M}NP}$ are
\begin{equation} \label{omega}
\omega^{-i}=-\omega^{i-}=\frac{1}{2}A_{ii} \, x^i\,dx^+ 
\end{equation}
From (\ref{forms})
\begin{equation}
\begin{split} \Gs&=2e^{i\alpha}\left(\Gamma^5+i\Gamma^6\right)
\left(\Gamma^7+i\Gamma^8\right)\Gamma^+ \\ \Fs&=-\frac{1}{\sqrt{2}}
\left(\Gamma^{1234}+\Gamma^{5678}\right)\Gamma^+ \end{split}
\end{equation} 
The equation involving the variation of the dilatino is easy to
solve and gives \begin{equation} \label{cond1}
\left(1+i\Gamma^{56}\right)\left(1+i\Gamma^{78}\right)
\left(1-\Gamma^{09}\right)\epsilon=0
\end{equation} 
If we represent the spinors in a basis of eigenvectors of the
Lorentz operators $\{\Gamma^{09},i\Gamma^{12},\newline
i\Gamma^{34},i\Gamma^{56},i\Gamma^{78}\}$, then a generic spinor can
be written as $(\pm,\pm,\pm,\pm,\pm)$. The solutions of (\ref{cond1})
are \begin{equation} \label{sol1}
\begin{array}{llll} (+,\pm,\pm,\pm,\pm) \qquad (1)\\ (-,\pm,\pm,-,-) 
\qquad (2)\\ (-,\pm,\pm,+,-)  \qquad (3)\\ (-,\pm,\pm,-,+) 
\qquad (4)\end{array}
\end{equation} 
The condition $\Gamma_{11}\epsilon=\epsilon$ imposes, moreover,
that there be an even number of ``$-$'' eigenvalues. Not all of these
solutions are supersymmetries of the background: the more involved
gravitino equation must also be satisfied. Because of
(\ref{susyeqns}), (\ref{omega}) and $\left(\Gamma^+\right)^2=0$ it is
found that \begin{equation}
\begin{array}{ll} \partial_-\epsilon=0 \\ 
\partial_i\epsilon=-i\Omega_i\epsilon-i\Lambda_i\epsilon \end{array}
\end{equation} where \begin{eqnarray} \Omega_i &\equiv&
-\frac{1}{4\sqrt{2}}\Gamma^+\lp\Gamma^{1234}+\Gamma^{5678}\rp\Gamma_i
\nonumber \\ \Lambda_i &\equiv& \frac{i}{8}e^{i\alpha}\lp
2\lp\Gamma^5+i\Gamma^6\rp\lp\Gamma^7+i\Gamma^8\rp\Gamma^+\Gamma_i+
\Gamma^i\lp\Gamma^5+i\Gamma^6\rp\lp\Gamma^7+i\Gamma^8\rp\Gamma^+\rp
\nonumber \end{eqnarray} 
Again because of $\left(\Gamma^+\right)^2=0$ we have
that $\Omega_i\Omega_j=\Omega_i\Lambda_j=\Lambda_i\Lambda_j=0$ for any
$i,j=1,\ldots,8$ and then $\partial_i\partial_j\epsilon=0$: $\epsilon$
can only depend linearly on the $x^i$'s

\begin{eqnarray} \label{epsilon}
\lefteqn{ \epsilon=
\chi-i\sum_{j=1}^{8}x^j\lp\Omega_j\chi+\Lambda_j\chi^*\rp={} }
\nonumber\\ &&{}=\lp
1-i\sum_{j=1}^{8}x^j\Omega_j\rp\chi-i\sum_{j=1}^{8}x^j\Lambda_j^*
\end{eqnarray}
where $\chi=\chi(x^+)$ is a positive chirality spinor to be determined
via the $\delta\psi_+=0$ equation and (\ref{cond1}). Substitution of
(\ref{epsilon}) into (\ref{cond1}) gives \begin{equation} \label{chi1}
\left(1+i\Gamma^{56}\right)\left(1+i\Gamma^{78}\right)
\left(1-\Gamma^{09}\right)\chi=0 \end{equation} thus $\chi$ will be
one of (\ref{sol1}).  Substituting (\ref{epsilon}) into
$\delta\psi_+=0$ and setting the constant term and the 8 terms linear
in $x^i$ separately equal to zero, we find that the following
equations must be solved \begin{equation}
\begin{split} &a)\quad\partial_+\chi =- \frac{i}{2\sqrt{2}}
\lp\Gamma^{1234}+\Gamma^{5678}\rp\Gamma^+\Gamma^-\chi+
\frac{1}{4}e^{i\alpha}\lp\Gamma^5+i\Gamma^6\rp\lp\Gamma^7+
i\Gamma^8\rp\lp1-\Gamma^+\Gamma^-\rp\chi^* \\
&b)\quad\Omega_j(\partial_+\chi) +
\Lambda_j(\partial_+\chi^*)+\frac{i}{2}A_{jj}\Gamma_j\Gamma^+
\chi-\frac{i}{2\sqrt{2}}\lp\Gamma^{1234}+
\Gamma^{5678}\rp\lp\Omega_j\chi+\Lambda_j\chi^\star\rp+
\\ &\qquad+
\frac{1}{2}e^{i\alpha}\lp\Gamma^5+i\Gamma^6\rp\lp\Gamma^7+
i\Gamma^8\rp\lp\Omega_j\chi^*+\Lambda_j^*\chi\rp=0
\end{split} \end{equation} 
We then substitute {\it a)} into {\it b)}, and find
that $\chi$ must satisfy \begin{eqnarray} \label{cond2}
\frac{1}{2}e^{i\alpha}\Omega_j\lp\Gamma^5+
i\Gamma^6\rp\lp\Gamma^7+i\Gamma^8\rp\chi^*+
\frac{1}{2}e^{-i\alpha}\Lambda_j\lp\Gamma^5-i\Gamma^6\rp\lp\Gamma^7-
i\Gamma^8\rp\chi+i\,A_{jj}\Gamma_j\Gamma^+\chi+ \nonumber \\
-\frac{i}{\sqrt{2}}\lp\Gamma^{1234}+\Gamma^{5678}\rp\lp\Omega_j\chi+
\Lambda_j\chi^*\rp+e^{i\alpha}\lp\Gamma^5+i\Gamma^6\rp\lp\Gamma^7+
i\Gamma^8\rp\lp\Omega_j\chi^*+\Lambda_j^*\chi\rp=0 \nonumber \\
\end{eqnarray} We notice that if $\Gamma^+\chi=0$, which corresponds
to the first of (\ref{sol1}), then equation (\ref{cond2}) is
satisfied: our solution has at least 16 supersymmetries \footnote{This
is a general result for type IIB superstrings on a pp--wave
\cite{Blau}.}. Let us now consider cases (2)-(4): after some
gamma--matrices algebra, we rewrite (\ref{cond2}) for $j=1,2,3,4$ as
\begin{eqnarray} \label{cond3} i \lp A_{jj}+\frac{1}{2} \rp
\Gamma^+\Gamma^j\chi= &-& \frac{i}{16}\Gp\Gamma^+\Gamma^j\Gm\chi+
\nonumber \\ &-& \frac{e^{i\alpha}}{4\sqrt{2}}\lp\Gamma^{1234}+
\Gamma^{5678}\rp\Gamma^+\Gamma^j\Gp\chi^* \end{eqnarray} The left--hand
side can never be zero, because we are considering the case in which
$\Gamma^+\chi\neq 0$, thus there can be a solution to this equation
only if the right--hand side of the equation doesn't vanish. Let us
consider the three cases (2)-(4) of (\ref{sol1}) \bd
\begin{array}{lllll} (2)\rightarrow \left\{\begin{array}{ll} 
\lp\Gamma^5-i\Gamma^6\rp \chi \neq 0 \\ \lp\Gamma^7-
i\Gamma^8\rp\chi\neq 0 \end{array} \right. \qquad \mathrm{and}\qquad 
\left\{\begin{array}{ll} \lp\Gamma^5+i\Gamma^6\rp \chi^* \neq 0 \\ 
\lp\Gamma^7+i\Gamma^8\rp\chi^*\neq 0 \end{array} \right. \\ \\
(3)\rightarrow \left\{\begin{array}{ll} \lp\Gamma^5-i\Gamma^6\rp \chi
= 0 \\ \lp\Gamma^7-i\Gamma^8\rp\chi\neq 0 \end{array} \right. \qquad
\mathrm{and} \qquad\left\{\begin{array}{ll} \lp\Gamma^5+i\Gamma^6\rp
\chi^* = 0 \\ \lp\Gamma^7+i\Gamma^8\rp\chi^*\neq 0 \end{array}
\right.\\ \\ (4)\rightarrow \left\{\begin{array}{ll}
\lp\Gamma^5-i\Gamma^6\rp \chi \neq 0 \\ \lp\Gamma^7-i\Gamma^8\rp\chi=
0 \end{array} \right. \qquad \mathrm{and}
\qquad\left\{\begin{array}{ll} \lp\Gamma^5+i\Gamma^6\rp \chi^* \neq 0
\\ \lp\Gamma^7+i\Gamma^8\rp\chi^*= 0 \end{array} \right.
\end{array}
\ed Thus we see that (3) and (4) cannot be solutions of
(\ref{cond3}). The only possible solution is (2). The first term on
the right--hand side of (\ref{cond3}) evaluated on
$\chi=(-,\pm,\pm,-,-)$ gives $i\,\Gamma^+\Gamma^j\chi$. As we already
mentioned the condition $\Gamma_{11}\chi=\chi$ implies that either one
of the two $\pm$'s must be a $-$ and the other a $+$. It follows then
that $\Gamma^{1234}\chi=\chi$, while $\Gamma^{5678}\chi=-\chi$ and
$\lp\Gamma^{1234}+\Gamma^{5678}\rp\chi=0$. The second term on the
right--hand side of (\ref{cond3}) becomes
$2\sqrt{2}e^{i\alpha}\Gamma^+\Gamma^j\Gamma^{57}\chi^*$, and equation
(\ref{cond3}) reads ($A_{jj}=1$ for $i,j=1,2,3,4$) \begin{equation}
\chi=-2\sqrt{2}\,i\,e^{i\alpha}\Gamma^{57}\chi^* \end{equation} which has
 no
solutions.


\end{document}